\documentclass[prl,twocolumn,twoside,showpacs,superscriptaddress]{revtex4}
\usepackage{graphicx,amssymb,amsmath,epsf,bm}
\epsfclipon

\newcommand{\rd}{\mathrm{d}}
\newcommand{\pv}{{\bm{v}}}
\newcommand{\expct}[1]{\langle #1 \rangle}

\begin{document}

\title{Lyapunov analysis captures the collective dynamics of large chaotic systems}


\author{Kazumasa A. Takeuchi}
\affiliation{Service de Physique de l'\'Etat Condens\'e,~CEA -- Saclay,~91191~Gif-sur-Yvette,~France}%
\affiliation{Department of Physics, The University of Tokyo, 7-3-1 Hongo, Tokyo 113-0033, Japan}%

\author{Francesco Ginelli}
\affiliation{Institut des Syst\`emes Complexes de Paris Ile-de-France, 57-59 Rue Lhomond, 75005 Paris, France}%
\affiliation{Service de Physique de l'\'Etat Condens\'e,~CEA -- Saclay,~91191~Gif-sur-Yvette,~France}%

\author{Hugues Chat\'e}
\affiliation{Service de Physique de l'\'Etat Condens\'e,~CEA -- Saclay,~91191~Gif-sur-Yvette,~France}%

\date{\today}

\begin{abstract}
We show, using generic globally-coupled systems, 
that the collective dynamics of large
chaotic systems is encoded in their Lyapunov spectra: 
most modes are typically localized
on a few degrees of freedom, but some are delocalized,
acting collectively on the trajectory.
For globally-coupled maps, we show moreover
a quantitative correspondence between the collective modes
and some of the so-called Perron-Frobenius dynamics.
Our results imply that the conventional definition
of extensivity must be changed as soon as collective dynamics sets in.
\end{abstract}


\maketitle

A common way of characterizing chaos is to measure
Lyapunov exponents (LE), which quantify
 the infinitesimal rate(s) of divergence of trajectories in phase space,
and to represent them arranged by decreasing order in a spectrum.
For large, spatially-extended, dissipative systems,
Ruelle conjectured that if Lyapunov spectra obtained at
different system sizes collapse onto a single curve when the exponent index is
rescaled by the system's volume, then chaos is extensive 
\cite{Ruelle-CommunMathPhys1982}.
This was indeed shown to hold for generic models of spatiotemporal chaos in
one space dimension  \cite{extensivity}.
However, in higher space dimensions or for globally-coupled systems,
 the extensivity of Lyapunov spectra may be questioned.
Indeed, it is now well known that such large chaotic systems,
in contrast to their one-dimensional counterparts,
 generically show non-trivial collective behavior,
 in which macroscopic observables evolve periodically, quasiperiodically,
 or even chaotically in time
 without exact synchronization of microscopic degrees of freedom
 \cite{Chate_Manneville-PTP1992, GloballyCouplingNTCB}. 
If such behavior is encoded in the Lyapunov spectrum, then the 
 above definition of extensivity cannot hold, because collective modes
are by definition intensive. Beyond this, knowing 
whether emerging macroscopic behavior can be captured by
traditional ``microscopic'' Lyapunov analysis (and if yes, how and to what extent)
is important for our general understanding of dynamical systems.

Few previous works approached this question, with contradicting conclusions:
For globally-coupled chaotic maps it was
 argued that one needs {\it finite-amplitude} macroscopic perturbations
 to quantify the instability of collective chaos
 \cite{Shibata_Kaneko-PRL1998, Cencini_etal-PhysD1999},
suggesting that traditional Lyapunov analysis cannot capture such macroscopic dynamics.
On the other hand, Nakagawa and Kuramoto \cite{Nakagawa_Kuramoto-PhysD1995},
 studying globally-coupled limit-cycle oscillators,
 have pointed at the possible connection
 between some LE and collective dynamics,
 although they did not offer any criterion to distinguish, or even define,
 such collective modes.

In this Letter, we present evidence that the collective dynamics of large
chaotic systems is encoded in their Lyapunov spectra in a rather simple 
manner: Whereas most modes collected in the spectrum are typically localized
on a few degrees of freedom, some specific modes are delocalized,
acting collectively on the trajectory. Our results rely on the investigation of the
covariant Lyapunov vectors (CLV) associated with the exponents.
Working, for simplicity, on globally-coupled systems, we show that
strong finite-size effects coupling microscopic and macroscopic modes
have to be overcome to unravel the underlying low-dimensional collective dynamics. 
For globally-coupled maps, these conclusions are strengthened by a 
direct study of the collective dynamics via the so-called
Perron-Frobenius (PF) operator: We show a
quantitative correspondence between some LE and CLV of the
PF dynamics and the collective modes present in the usual Lyapunov analysis.
We finally discuss how our results imply that the conventional definition
of extensivity must be changed as soon as collective dynamics sets in.

Covariant Lyapunov vectors span the subspaces of the Oseledec
decomposition of tangent dynamics, and thus provide intrinsic directions of growth of
perturbations for each LE $\lambda^{(j)}$
\cite{Eckmann_Ruelle-RMP1985}.
It is only recently that CLV became numerically accessible 
for large systems \cite{Ginelli_etal-PRL2007} and they must not be confused
with the Gram-Schmidt vectors customarily used when calculating LEs, which
are not intrinsic and usually bear no physical meaning.
The average localization of CLV  $\pv^{(j)}$
can be quantified by calculating the so-called inverse participation ratio
 \cite{Mirlin-PhysRep2000}
$Y_2^{(j)} \equiv \langle{\sum_i |v_i^{(j)}|^4}\rangle_t$,
where the brackets indicate averaging along the trajectory.
(Here, the CLV are normalized with the L2 norm $\sum_i |v_i^{(j)}|^2=1$.)
Since $Y_2^{(j)}$ is just the inverse average number of degrees of freedom 
participating in $\pv^{(j)}$,
collective (delocalized) and microscopic (localized) modes are defined
as those satisfying, respectively, $Y_2 \sim 1/N$ and $Y_2 \sim \text{const.}$
in the $N\to\infty$ limit.

\begin{figure}[t]
  \includegraphics[clip,width=8.6cm]{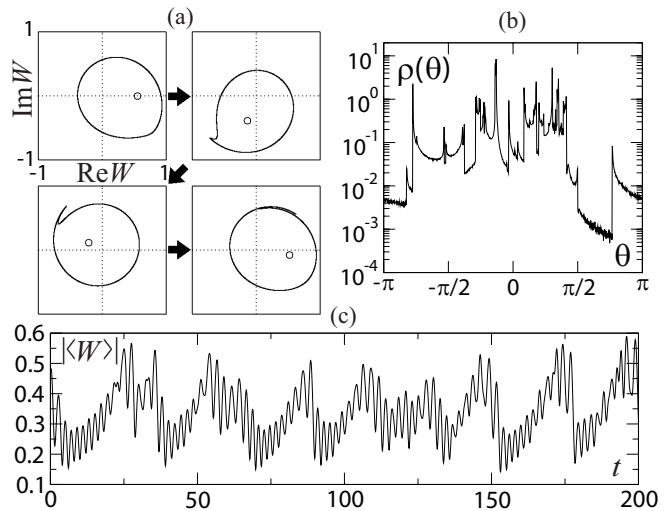}
  \caption{Collective chaos in globally-coupled 
limit-cycle oscillators [Eq.\ \eqref{eq:GLDef}] with $c_1 = -2.0$, $c_2 = 3.0$, $K = 0.47$, $N = 10^7$. (a) Typical snapshots of the population in the complex plane; 
note the stretching and folding of the supporting line. 
(b) Typical distribution $\rho(\theta \equiv \arg W)$
 for a configuration where no fold is present.
(c) Time series $|\langle W\rangle|$.}
  \label{fig:SFState}
\end{figure}

To start, we consider,
 following \cite{Nakagawa_Kuramoto-PhysD1995}, a simple system of $N$
globally-coupled limit-cycle oscillators known for exhibiting non-trivial
collective behavior:
\begin{equation}
 \dot{W}_i = W_i - (1+\mathrm{i}c_2)|W_i|^2 W_i + K (1+\mathrm{i}c_1) (\expct{W} - W_i),  \label{eq:GLDef}
\end{equation}
 where $W_i$ are complex variables and $\expct{W} \equiv \frac{1}{N}\sum_i W_i$.
Here we focus on a collective chaos regime where individual oscillators, 
while behaving erratically, arrange themselves, in the complex plane, on a curve undergoing
stretching and folding along time (Fig.\ \ref{fig:SFState}a).
The density of oscillators along this fractal-like curve is a complicated function with many peaks,
reminiscent of the invariant measure of single chaotic maps (Fig.\ \ref{fig:SFState}b).
In this regime, the evolution of collective variables such as $\expct{W}$ takes 
the form of a weakly chaotic modulation of some quasiperiodic signal
(Fig.\ \ref{fig:SFState}c). 
This type of dynamics is similar to
 some small systems, 
made of, e.g., a couple of nonlinear oscillators with incommensurate 
frequencies
 \cite{Sano_Sawada-PhysLett1983}.

\begin{figure}[t]
  \includegraphics[clip,width=8.6cm]{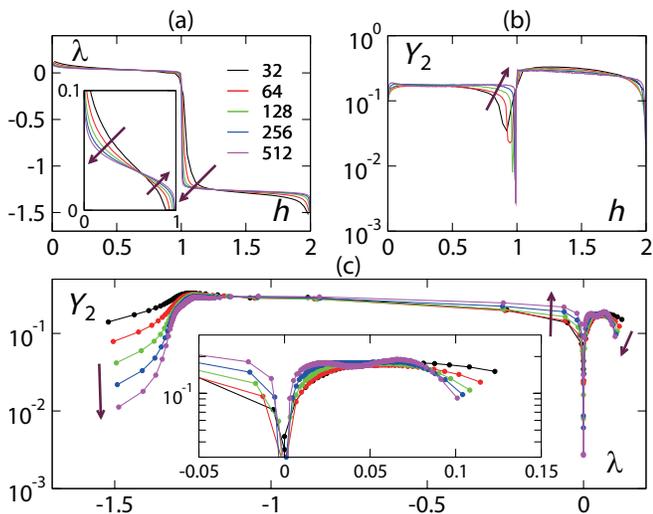}
  \caption{(color online) 
Spectra of $\lambda$ vs. $h$ (a), $Y_2$ vs. $h$ (b), and $Y_2$ vs. $\lambda$ (c) 
for the system depicted in Fig.~\ref{fig:SFState} for different sizes $N$. Insets: close-ups of the positive branch.
(Arrows indicate increasing $N$.)}
  \label{fig:Spectra}
\end{figure}

We calculated LE and the associated CLV
 for different system sizes (Fig.~\ref{fig:Spectra}),
 using the algorithm described in \cite{Ginelli_etal-PRL2007}.
Lyapunov spectra, plotted as functions of the rescaled index
$h \equiv (j-0.5)/N$, are composed of two main branches
 near $0.05$ or near $-1.25$,
 reflecting the basic dynamics of individual (uncoupled) oscillators
 (Fig.~\ref{fig:Spectra}a).
Both branches show a systematic drift as increasing $N$
 and thus the Lyapunov spectra
 do {\it not} collapse entirely onto a single curve
 (inset of Fig.~\ref{fig:Spectra}a).
Similarly, the averaged participation ratios $Y_2$ of the
CLV form two groups largely independent of $N$, 
corresponding hence to localized vectors,
 except the edges of the two groups (Fig.~\ref{fig:Spectra}b).
Indeed, the parametric plots $Y_2^{(j)}$ vs  $\lambda^{(j)}$
 reveal that at both ends 
of the spectra, as well as near zero exponents,
 $Y_2$ decreases with $N$, suggesting 
the possible presence of delocalized modes (Fig.~\ref{fig:Spectra}c). 

\begin{figure}[t]
  \includegraphics[clip,width=8.6cm]{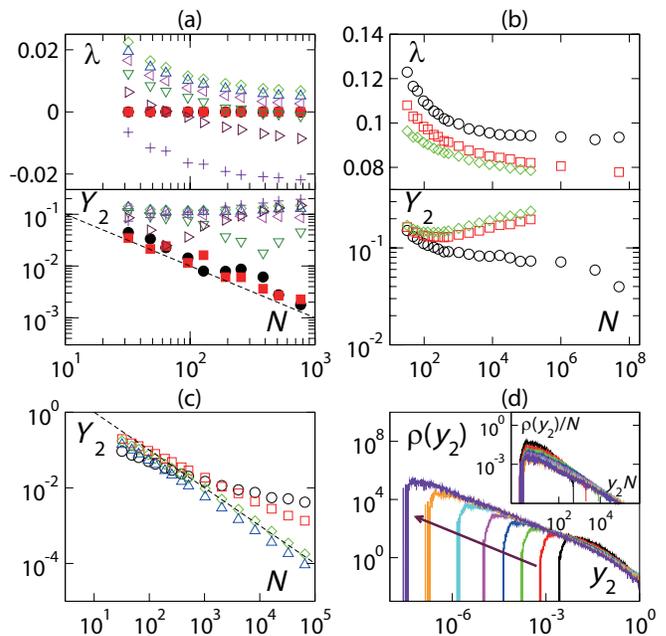}
  \caption{(color online) Size dependence for the system depicted in 
Figs.\ \ref{fig:SFState} and \ref{fig:Spectra}.
(a) $\lambda$ and $Y_2$ vs $N$ around $\lambda=0$. The two modes with $\lambda=0$ 
are shown in filled symbols.
(b) same as (a) for the first 3 modes.
(c) $Y_2$ vs $N$ for the last four modes 
(d) Distribution of instantaneous $Y_2$ values for the first mode vs $N$. Inset:
rescaling showing that eventually $Y_2\propto 1/N$.
[The dashed lines in (a,c) correspond to the power-law $Y_2 \sim 1/N$.]}
  \label{fig:LyapPtcpVsN}
\end{figure}

Focusing first on near-zero exponents reveals the existence
of {\it two} numerically-null exponents ($\lambda\sim10^{-5}$ at our numerical resolution) 
with $Y_2\propto1/N$, {\it i.e.} delocalized modes 
(solid symbols in Fig.\ \ref{fig:LyapPtcpVsN}a). 
Nearby exponents ``cross'' the zero line smoothly as $N$ is varied, 
and their $Y_2$ are essentially independent of $N$, except when they come accidentally 
close to the two collective zeros, in which case the algorithm cannot 
resolve well this degeneracy.
Among the most negative exponents, only the last two appear delocalized when 
large-enough system sizes are considered (Fig.\ \ref{fig:LyapPtcpVsN}c) \cite{NOTE}. 
Finally, one has to explore even larger system sizes to see that the first mode is actually
delocalized: for $N>10^6$, $Y_2$ starts decreasing faster (Fig.\ \ref{fig:LyapPtcpVsN}b).
The scaling of left part of the distribution of instantaneous $Y_2$ values indicates that
asymptotically $Y_2\propto 1/N$  (Fig.\ \ref{fig:LyapPtcpVsN}d).

The participation ratio $Y_2$ is only a global indicator
which does not provide information about the actual structure 
of the CLV. 
We have investigated this structure in a careful study. While details
will be published elsewhere\cite{Takeuchi_etal-TBP}, 
we only report here the main findings.
Microscopic, localized, modes, are each localized on two nearby oscillators.
The vector of the chaotic collective mode
typically moves some of the peaks in the oscillator 
density along the curve on which it is located in the complex plane. 
Different peaks are moved at different times, increasing thus 
the global disorder.
On the other hand, the vectors of the two delocalized negative modes tend
to adjust the width of these peaks, increasing synchronization. 
As for the two collective zero modes, their vectors are 
difficult to identify due to the degeneracy of the exponents, but they can be
assigned to a global change of phase and to a translation along their trajectory 
of all oscillators, two ``natural'' neutral modes.

To sum up the set of collective LE correspond to what one would expect
from the observed global dynamics in Fig.\ \ref{fig:SFState}c, and the action
of the associated CLV also corroborates this view.
Although strong finite-size effects couple macroscopic (delocalized) and microscopic (localized)
modes, it is likely that, for the case studied above, no other collective mode exists,
although we cannot exclude the emergence of other ones 
(in particular very weakly chaotic) at still larger sizes than those probed here.
Thus, conventional Lyapunov analysis, supplemented with the calculation of covariant
vectors, is able to capture the collective behavior of a large chaotic system.
To investigate the generality of this finding, we now turn to systems of globally-coupled maps of
one real variable.

In globally-coupled systems of identical units, 
one can look ``directly'' at the infinite-size limit by
studying the evolution of the instantaneous distribution function of the values 
taken by the units, which is governed by the Perron-Frobenius operator.
Although this is a difficult analytical task, a numerical approach,
e.g., by using finer and finer binnings of the support of 
the evolving distribution, is in principle possible.
In the case described above, this is made difficult by the fact that this 
support is the complex plane fractal-like curve described
in Fig.~\ref{fig:SFState}a. 
For globally-coupled maps of a real variable, however, the support is
typically a bounded interval of the real axis, so that a numerical integration is
accessible. As a matter of fact,
previous work on globally-coupled noisy logistic maps
 \cite{DeMonte_etal,Shibata_etal-PRL1999} 
has provided partial evidence 
that the LE of the PF dynamics are related to the collective dynamics, 
but no direct link was ever shown.
We thus consider the same system:
\begin{equation}
 x_i^{t+1} = (1-K)f(x_i^t) + K \expct{f(x)} + \xi_i^t,  \label{eq:GCMDef}
\end{equation}
 where $f(x) \equiv 1-ax^2$ is the logistic map
 (here we use $a=1.57$, in the one-band chaos regime)
 and $\xi_i^t$ is a delta-correlated noise.
For an infinite system, the evolution of $\rho^t(x)$, the instantaneous 
distribution of $x$ values, is governed by the following nonlinear PF equation
\begin{equation}
\begin{array}{ll}
 & \rho^{t+1}(x) = \int \rho_{\rm N}(F^t(y) - x) \rho^t(y) \rd y\\
{\rm with} &
F^t(y) = (1-K)f(y) + K \int f(z)\rho^t(z) \rd z \;,  
\label{eq:PFDef}
\end{array}
\end{equation}
 where $\rho_{\rm N}(\xi)$ is the noise distribution function.
To evolve properly both  Eq.~\eqref{eq:PFDef} and its tangent space dynamics 
requires to keep the support of  $\rho^t(x)$ within $[-1,1]$, 
the invariant interval of the logistic map,
and to have well-defined derivatives of  $\rho_{\rm N}(\xi)$, a fact overlooked by 
 \cite{DeMonte_etal,Shibata_etal-PRL1999}. We therefore use
the bounded and differentiable noise Kumaraswamy distribution
 $\rho_{\rm N}(\xi) = 15{\xi'}^2(1-{\xi'}^3)^4$
 with $\xi' \equiv (\xi/\sigma+1)/2 \in [0,1]$,
so that  $\xi \in [-\sigma,\sigma]$.
We focus on the collective chaos regime observed
 for $K=0.28$ and a noise level $\sigma=0.1$,
 in which local dynamical variables tend to synchronize
 but are weakly scattered by microscopic chaos and noise
 \cite{Teramae_Kuramoto-PRE2001, DeMonte_etal}.

\begin{figure}[t]
  \includegraphics[clip,width=8.6cm]{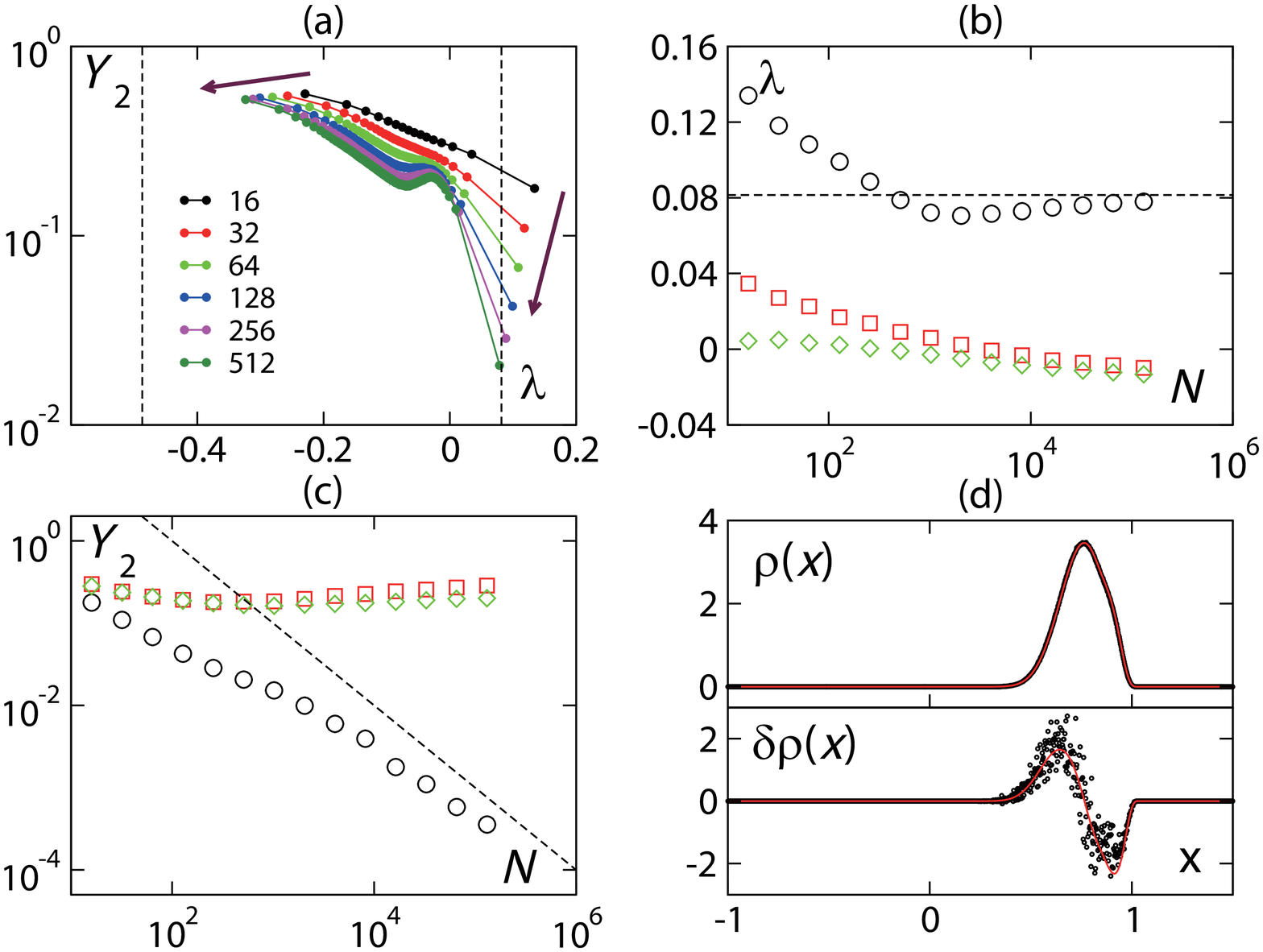}
  \caption{(color online) Lyapunov analysis for globally-coupled 
noisy logistic maps [Eq.\ \eqref{eq:GCMDef}] with $a=1.57, K=0.28, \sigma=0.1$ 
and the corresponding PF equation [Eq.\ \eqref{eq:PFDef}]. 
(a) $(Y_2,\lambda)$ spectra for different $N$. The vertical lines 
indicate the location of the first two LE of the PF dynamics. 
(b) First three LE vs system size $N$ (symbols) and for first PF LE 
(dashed line). 
(c) $Y_2$ vs $N$ for the first three modes [same symbols as in (b), 
the dashed line indicates the power law $Y_2 \sim 1/N$]. 
(d) Typical snapshot for the distribution $\rho(x)$ (top) and for the first CLV $\delta\rho(x)$ (bottom). Solid lines and circles are the results from PF and maps of size $N=10^7$, respectively.}
\label{fig:NoisyLogisticGCM}
\end{figure}

We first describe the outcome of the conventional Lyapunov analysis.
The $(\lambda,Y2)$ spectrum of small-size systems (Fig.\ \ref{fig:NoisyLogisticGCM}a)
reveals that the first few (positive) modes show signs of being delocalized. 
Calculating only the first three in large systems shows that the first mode is indeed
delocalized, while the following ones are localized  
(Fig.\ \ref{fig:NoisyLogisticGCM}c). Similar calculations 
do not indicate that any of the last modes is actually delocalized (data not shown). 
Thus, conventional Lyapunov analysis of finite systems seems to detect the 
presence of only one collective mode with exponent $\lambda_{N\to\infty}^{(1)}\simeq 0.081$
 (Fig.\ \ref{fig:NoisyLogisticGCM}b).

We now turn to our investigation of the PF dynamics.
The evolution of $\rho(x)$ matches that of large collections of maps
 for a large number of timesteps
 (not infinite, due to finite-size and finite-binning effects),
 when taking for initial condition of the PF dynamics
 a (smoothed) instantaneous distribution of maps.
The Lyapunov spectrum of the PF dynamics contains only 
one positive exponent $\lambda_{\rm PF}^{(1)}\simeq 0.0815$, in remarkable
agreement with the ``finite-size'' collective exponent 
$\lambda_{N\to\infty}^{(1)}$ calculated above (Fig.\ \ref{fig:NoisyLogisticGCM}b).
Moreover, the CLV associated with $\lambda_{N\to\infty}^{(1)}$ and
 $\lambda_{\rm PF}^{(1)}$ share the same structure.
In Fig.\ \ref{fig:NoisyLogisticGCM}d, we show snapshots of each vector at timesteps 
carefully chosen so that the two distributions ($\rho(x)$ from the PF dynamics and
the distribution reconstructed from a large-$N$ simulation) coincide very well (top panel).
Both vectors exert the same shift in distribution $\delta\rho(x)$ (lower panel).
Thus, the collective mode appearing in the conventional Lyapunov analysis
is the first Lyapunov mode of the PF dynamics.
However, the PF dynamics possesses infinitely-many modes, 
all negative except the first one in our case,
 and even the second one, given at  
$\lambda_{\rm PF}^{(2)}\simeq -0.49$, is not 
detected by conventional Lyapunov analysis,
 at least up to $N=131072$.
In fact, PF modes with \textit{negative} LE
may not necessarily be present in the microscopic Lyapunov analysis.
This is trivially true in the special case $K=0$ (no coupling),
 where all conventional Lyapunov modes are localized with the same, positive exponent,
 while PF dynamics shows infinitely many negative exponents which can be seen as
 superpositions of $O(N)$ independent microscopic modes.
Whether this zero-coupling situation extends to coupled maps
---and thus would explain how some PF negative modes could remain unseen 
in conventional spectra---
is a difficult question which should be treated at a more 
mathematical level.
We note finally that at least {\it some} negative collective modes appear in conventional 
Lyapunov analysis, as shown for Eq.\ \eqref{eq:GLDef},
 while positive PF modes are probably all present 
because they cannot be superpositions of independent
microscopic modes. (Such perturbations would not grow.)

To summarize, we have shown, using generic globally-coupled systems, that
conventional Lyapunov analysis can capture the collective dynamics 
of large chaotic systems through delocalization of CLV.
We have found collective Lyapunov modes which are delocalized
for large but finite system sizes, and clearly related 
to the observed macroscopic dynamics.
This implies that there is no general need for finite-amplitude perturbations
 as claimed in some earlier studies
 \cite{Shibata_Kaneko-PRL1998, Cencini_etal-PhysD1999}.
Moreover, we have directly identified, in one case at least, 
one such collective mode with a Lyapunov mode of the PF dynamics.
Further results will be needed to strengthen these findings. 
We have studied other globally-coupled systems and reached similar conclusions
\cite{Takeuchi_etal-TBP}, but
the strong ``finite-size'' coupling between microscopic and collective modes reported
here makes it difficult to build a reliable asymptotic picture. 
We believe our conclusions should also hold for locally-coupled systems, which generically
exhibit non-chaotic collective behavior, but first attempts in this direction 
have revealed even stronger size effects in this case.

At any rate,
 our results already imply that the conventional criterion usually adopted 
 to prove the extensivity of (microscopic) chaos in large dynamical systems
 has to be changed.
Collective modes should be excluded from the 
 usual rescaling of Lyapunov
 spectra at least,
 but this may not be enough in practice,
 since microscopic modes are strongly influenced
 by neighboring collective modes
 through the finite-size effects reported above.
Thus, an operative definition of chaos extensivity using Lyapunov analysis
 implies a quantitative understanding of these finite-size effects,
 a task left for future work.

We acknowledge fruitful discussions with A. Politi and M. Sano.
This work is supported in part by JSPS.


\begin{thebibliography}{99}
\bibitem{NOTE} One is quickly limited by numerical power when calculating the 
full spectrum of large systems,
but the first few LE and CLV
are accessible for larger sizes.
For globally-coupled systems, this is also true of the last few modes.
\end{thebibliography}
\end{document}